\documentclass[preprint,12pt,numbers,sort]{elsarticle}
\usepackage{graphicx}
\usepackage{epsfig}
\usepackage{epstopdf}
\usepackage{dcolumn}
\usepackage{bm}
\usepackage{amsmath}
\usepackage{color}
\usepackage{amssymb}

\usepackage[utf8]{inputenc}
\usepackage[T1]{fontenc}
\usepackage{indentfirst}

\usepackage{enumitem}
\usepackage{xspace}

\begin{document}


\begin{frontmatter}

\title{Symmetry restoration in the mean-field description of proton-neutron pairing}

\author{A.M. Romero$^a$, J. Dobaczewski$^{a,b,c}$, A. Pastore$^a$}
\address{
$^a$Department of Physics, University of York, Heslington, York YO10 5DD, United Kingdom \\
$^b$Institute of Theoretical Physics, Faculty of Physics, University of Warsaw, ul. Pasteura 5, PL-02-093 Warsaw, Poland \\
$^c$Helsinki Institute of Physics, P.O. Box 64, FI-00014 University of Helsinki, Finland}

\date{\today}

\begin{abstract}
We show that the symmetry-restored paired mean-field states
(quasiparticle vacua) properly account for isoscalar versus isovector
nuclear pairing properties. Full particle-number, spin, and isospin
symmetries are restored in a simple SO(8) proton-neutron pairing
model, and prospects to implement a similar approach in a realistic
setting are delineated. Our results show that, provided all symmetries
are restored, the pictures based on pair-condensate and
quartet-condensate wave functions represent equivalent ways of
looking at the physics of nuclear proton-neutron pairing.
\end{abstract}



\end{frontmatter}


A key question in nuclear structure physics is do proton-neutron (pn) pairs form collective condensates in
nuclei in the same way that like-particle pairs do?
Ever since the existence of
like-particle nuclear pairing was suggested in 1958 by Bohr,
Mottelson, and Pines~\cite{(Boh58)}, this simple question has
been addressed in numerous studies~\cite{(Fra14)}. As late as in 2004, the authors
of Ref.~\cite{(Per04a)} concluded that {\it in spite of many attempts
to extend the quasiparticle approach to incorporate the effect of pn
correlations, no symmetry-unrestricted mean-field calculations of pn
pairing, based on realistic effective interaction and the isospin
conserving formalism have been carried out}. This conclusion still holds even today.

In this Letter, we show that sometimes contradicting conclusions
about the existence of the pn pair condensate may have resulted from
using a mean-field formalism without full symmetry restoration. Here we apply
this formalism within simultaneous breaking and
then restoration of three major symmetries: particle-number,
angular-momentum, and isospin. In the shell-model framework these
symmetries are not broken and hence do not have to be restored. A
number of such studies already exist, see, e.g., Ref.~\cite{(Lei11)}.
However, the shell-model interprets the pn pairing as an
effect of a strong nucleon-nucleon isoscalar interaction, and is less
concerned with the analysis of wave functions in terms of collective condensates.
In this sense, the question of existence of
the putative pn condensate remains open.

Due to the attractive nature of the nuclear interaction, atomic
nuclei are strongly correlated systems exhibiting superfluid
properties. The theoretical description of nuclear superfluidity is
directly related to the theory of electronic superconductivity,
wherein Cooper pairs of electrons in time-reversed states condensate
near the Fermi level. In the nuclear case, we may expect a
possible formation of six types of pairs, corresponding to the four
degrees of freedom of the nucleon: spin and isospin, up and down.
More precisely, we may have scalar-isovector Cooper pairs
$\hat{P}^+_{\nu}$, with three projections of the total isospin
$\nu\equiv{}T_z$=$0,\pm1$, and vector-isoscalar pairs $\hat{D}^+_{\mu}$,
with three projections of the total spin $\mu\equiv{}S_z$=$0,\pm1$. The
condensation of spin-aligned $\hat{D}^+_{\mu}$ pairs has recently attracted
increased attention, see Refs.~\cite{(Isa16),(Kim18)} and
references cited therein.

The most general pair condensate is represented by a quasiparticle
vacuum. This can be written in terms of the Thouless state~\cite{(Tho60a),(Rin80a)},
which may be expressed as $|\Phi\rangle={\cal{N}}\exp\{\hat{Z}^+\}|0\rangle$,
for the Thouless pair $\hat{Z}^+$ given by
\begin{equation}\label{eq:thoupair}
\hat{Z}^+ = \sum_{\nu=0,\pm1}p_{\nu}\hat{P}^+_{\nu}+\sum_{\mu=0,\pm1}d_{\mu}\hat{D}^+_{\mu}.
\end{equation}
In the above equation, $p_{\nu}$ and $d_{\nu}$ are complex isovector and isoscalar amplitudes, respectively,
$|0\rangle$ is the particle vacuum and ${\cal{N}}$ is the normalization constant.

It is now obvious that in the Thouless state all symmetries:
particle-number, spin, and isospin, are strongly mixed. Therefore,
the standard paired-mean-field minimization of the average energy,
which in nuclear phy\-sics is called Hartree-Fock-Bogolyubov (HFB)
theory~\cite{(Rin80a)}, may or may not give the best result. A great
number of studies based on the HFB approach already exist, see,
e.g., Refs.~\cite{(Goo99),(Sat01d),(Sat01e),(Glo04a),(Gez11)} and reviews in
Refs.~\cite{(Per04a),(Fra14)}. In this Letter, we argue that in order to
analyze the problem of the pn pairing it is necessary to employ a more
sophisticated approach that is based on the minimization of energy
after all symmetries are restored.

The relevant method corresponds to the so-called variation-after-projection (VAP)~\cite{(Rin80a)} method,
which employs the projected Thouless states,
\begin{equation}\label{eq:thoustate2}
|\Phi^{AST}_{MK,NL}\rangle = \hat{P}_A \hat{P}^S_{MK}\hat{P}^T_{NL}|\Phi\rangle,
\end{equation}
as variational trial states. The projection operators: $\hat{P}_A$ on
particle number $A$, $\hat{P}^S_{MK}$ on total spin $S$ and its
projection $M$, and $\hat{P}^T_{NL}$ on total isospin $T$ and its
projection $N$, involve one-dimensional integration
over the gauge angle, three-dimensional integration over the
spin-rotation Euler angles and three-dimensional integration over
the isospin-rotation Euler angles~\cite{(Rin80a),(She18)}, respectively.
In this Letter, we report on the implementation of a complete
seven-dimensional integration which allows us to fully restore all
relevant symmetries that are broken in an arbitrary
symmetry-unrestricted Thouless state. Although such a technology has
already been previously applied in the shell-model
context~\cite{(Gao15a)}, below we argue that it is essential for
analyzing the physics of pn pairing.

However, before embarking on full-scale VAP calculations in a
realistic nuclear DFT setting, one would like to know if such a
complete and demanding approach is capable of bringing better
solutions when applied in a simple model. For that, in this Letter we
perform a full VAP analysis of the well-known SO(8)
model~\cite{(Flo64b),(Pan69a),(Eva81),(Eng97),(Dob98a),(Kot06),(Fra14)}.
To make the properties of the model as clear as possible, we begin by
a novel discussion of its building blocks and symmetries, and later
we recall its Hamiltonian and dynamics.

The building blocks of the model are the isovector and isoscalar
pairs within a single-particle phase space of a few degenerate $\ell$
shells,
\begin{eqnarray}\label{eq:pair-isovector}
\hat{P}^+_{\nu} &=& \sum_{\ell}\sqrt{\tfrac{2{\ell}+1}{2}} \left( a_{{\ell}\frac{1}{2}\frac{1}{2}}^+ a_{{\ell}\frac{1}{2}\frac{1}{2}}^+ \right)_{M=0,S_z=0,T_z=\nu}^{L=0,S=0,T=1},
\\ \label{eq:pair-isoscalar}
\hat{D}^+_{\mu} &=& \sum_{\ell}\sqrt{\tfrac{2{\ell}+1}{2}} \left( a_{{\ell}\frac{1}{2}\frac{1}{2}}^+ a_{{\ell}\frac{1}{2}\frac{1}{2}}^+ \right)_{M=0,S_z=\mu,T_z=0}^{L=0,S=1,T=0},
\end{eqnarray}
where $a_{{\ell}\frac{1}{2}\frac{1}{2}}^+$ are the creation operators of
a particle with orbital angular momentum ${\ell}$, spin $\frac{1}{2}$, and isospin
$\frac{1}{2}$. The round brackets denote triple standard Clebsch-Gordan
coupling to the total orbital angular momentum $L$, spin $S$, and isospin
$T$, having, respectively, projections $M$, $S_z$, and $T_z$.
The maximum number of particles allowed in this phase space
is equal to $4\Omega$ for $\Omega = \sum_{\ell}(2{\ell}+1)$. For
deformed nuclei with spin-orbit coupling taken into account, the
notion of spin should, in fact, be understood as that of the
alispin~\cite{(Sch10b)}, which pertains to a pair of deformed
Kramers-degenerate single-particle states.

In the past, much of the discussion related to properties of pn
pairing concentrated on the question of whether real or complex
quasiparticle amplitudes have to be used. In order to solve this
problem in the context of the most general Thouless pairs, defined in
Eq.~(\ref{eq:thoupair}), we briefly touch upon their symmetries. To
begin, let us consider a system described by a scalar and isoscalar
Hamiltonian, such as that in the SO(8) model, which is defined below.

In the first instance, we note that unabridged spin and isospin
projections involve rotating the Thouless pair over the full spin and
isospin SO(3) groups, hence, we can freely choose its initial
orientations in the spin and isospin spaces, respectively.
This means that
the vector and isovector pair-creation operators,
Eqs.~(\ref{eq:pair-isoscalar}) and (\ref{eq:pair-isovector}), can be
arbitrarily aligned along one of the directions in space and
isospace, respectively. Without any loss of generality, we can
choose orientations along the $z$ axes, that is, we can keep in
Eq.~(\ref{eq:thoupair}) only spherical amplitudes $p_0$ and $d_0$.
Then, the Thouless states become eigenstates of spin and isospin
projections with $S_z=T_z=0$. Such a choice has an enormous
advantage, namely, it allows for reducing the integrations over the
spin and isospin Euler angles to one dimension only, which
reduces seven dimensions of integration to just three. We have also
been able to test and benchmark all of our results by performing
unrestricted integrations.

Second, we note that by a simple expansion of the exponential
function, the particle-number projection of the Thouless state
$|\Phi\rangle={\cal{N}}\exp\{\hat{Z}^+\}|0\rangle$ is equal to
$|\Phi_A\rangle={\cal N}'(\hat{Z}^+)^{A/2}|0\rangle$ for
${\cal N}={\cal N}'(A/2)!$.
Therefore,
an overall multiplicative factor of the Thouless pair, and its
phase, can be absorbed in the normalization constant ${\cal N}'$,
and are thus irrelevant. This allows us to parametrize the
most general Thouless pair expressed in Eq.~(\ref{eq:thoupair}) in terms of
two angles $0\leq\alpha<\pi$ and $0\leq\varphi<\pi$ only, that is,
$\hat{Z}^+(p_0,d_0)\equiv\hat{Z}^+(\alpha,\varphi)$ for
\begin{equation}\label{eq:parametrization}
p_0 = \sin(\tfrac{1}{2}\alpha)e^{-i\varphi} \quad\mbox{and}\quad
d_0 = \cos(\tfrac{1}{2}\alpha)e^{ i\varphi}.
\end{equation}
The angle $\alpha$ ($\varphi$) controls the relative amplitude (phase)
between the isovector and isoscalar pairs.

Third, we have to take into account the fact that every scalar and isoscalar
Hamiltonian is also invariant with respect to the spin and isospin
signatures, $\hat{\cal{S}}\equiv\exp(i\pi\hat{S}_y)=i\hat{\sigma}_y$
and $\hat{\cal{T}}\equiv\exp(i\pi\hat{T}_y)=i\hat{\tau}_y$, respectively, which
rotate spins and isospins by angle $\pi$ about the
corresponding $y$ axes.
Transformation rules of the isovector and isoscalar pairs under such
rotations follow directly from the general rules of how scalars and
vectors are transformed. Indeed, any scalar is invariant with respect
to the rotation by angle $\pi$, and any vector then changes sign. It
thus follows that the scalar pairs are
${\cal{S}}$-even-${\cal{T}}$-odd and the isoscalar pairs are
${\cal{S}}$-odd-${\cal{T}}$-even, and thus the Thouless pairs
(\ref{eq:thoupair}) transform as:
\begin{eqnarray}
\label{eq:S-transform}
\hat{\cal{S}}\hat{Z}^+(\alpha,\varphi)\hat{\cal{S}}^+ &=&i\hat{Z}^+(\alpha,\varphi+\tfrac{\pi}{2}),
\\ \label{eq:I-transform}
\hat{\cal{T}}\hat{Z}^+(\alpha,\varphi)\hat{\cal{T}}^+ &=&i\hat{Z}^+(\alpha,\varphi-\tfrac{\pi}{2}).
\end{eqnarray}

Finally, we have to fix the phase convention. Here we adopt the one of
Condon-Shortley in the LS basis, by which all single-particle states
transform under time-reversal $\hat{T}$ as,
\begin{equation}\label{eq:convention}
\hat{T}a_{{\ell}m;\frac{1}{2}s_z;\frac{1}{2}t_z}^+\hat{T}^+
=(-1)^{\ell+m}(-1)^{\frac{1}{2}+s_z}a_{{\ell}-m;\frac{1}{2}-s_z;\frac{1}{2}t_z}^+.
\end{equation}
This convention carries over to the isovector and isoscalar pairs,
Eqs.~(\ref{eq:pair-isovector}) and  (\ref{eq:pair-isoscalar}),
which turn out to be time-even and time-odd, respectively.
As a consequence, the Thouless pairs transform under time reversal as
\begin{equation}\label{eq:T-transform}
\hat{T}\hat{Z}^+(\alpha,\varphi)\hat{T}^+ =i\hat{Z}^+(\alpha,\tfrac{\pi}{2}-\varphi).
\end{equation}

Altogether, we see that the invariance of the Hamiltonian with
respect to the spin or isospin signature renders the average energies
periodic in $\varphi$ with period of $\tfrac{\pi}{2}$, whereas that
with respect to the time reversal renders them symmetric with respect
to the line at $\varphi=\tfrac{\pi}{4}$. At this line, the Thouless
pairs are time-even (up to an irrelevant phase factor).

Since our entire analysis of symmetries is performed for the Thouless
states, we avoid any possible ambiguities related to definitions and
phase conventions of quasiparticle states, density matrices, and
pairing tensors, which can be now consistently determined from the
Thouless pairs using generic expressions~\cite{(Rin80a)}.

\begin{figure*}[!t]
\begin{center}
\includegraphics[width=\textwidth]{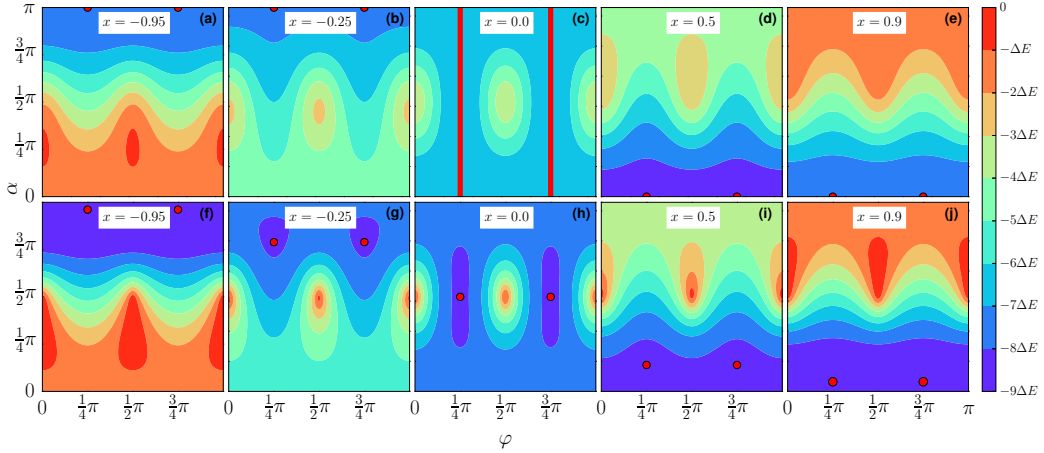}
\end{center}
\caption{Average values of the SO(8) Hamiltonian
(\protect\ref{eq:ham}), calculated for the unprojected (upper panels)
and projected (\ref{eq:thoustate2}) (lower panels) Thouless states,
parametrized by angles $\alpha$ and $\varphi$ as in
Eqs.~(\protect\ref{eq:thoupair}) and
(\protect\ref{eq:parametrization}). Calculations were performed for
$\Omega=12$, with projection on $A=24$ and $T=S=0$. From left
to right panel show results for $x=-0.95$, $-$0.25, 0, 0.5, and 0.9,
and color bands correspond to steps of $\Delta{E}=20$, 15, 13, 17,
and 20, respectively. All results are in units of $g$.
}
\label{Fig:maps}
\end{figure*}

For the Hamiltonian of the SO(8) model we use the representation
introduced in Ref.~\cite{(Kot06)},
\begin{equation}\label{eq:ham}
\hat{H} = - g(1-x) \sum_{\nu=0,\pm1} \hat{P}^+_{\nu} \hat{P}_{\nu}
          - g(1+x) \sum_{\mu=0,\pm1} \hat{D}^+_{\mu} \hat{D}_{\mu}.
\end{equation}
The model makes it possible to study the dynamical properties and relative importance
of the isoscalar and isovector modes of pairing.
Indeed, with the overall pairing strength controlled by parameter $g$,
the relative importance of the isovector vs.~isoscalar pairing is governed by the mixing
parameter $x$. For $x=+1(-1)$, the Hamiltonian has a pure isoscalar
(isovector) character, whereas within the interval $-1<x<1$, we should
expect a competition between the two possible types of pairing.
Using group-theory methods the Hamiltonian, specified by Eq.~(\ref{eq:ham}), can be
diagonalized exactly~\cite{(Pan69a),(Kot06)}.

In Fig.~\ref{Fig:maps}, we show average values of the SO(8)
Hamiltonian (\ref{eq:ham}) calculated for the unprojected Thouless
states (upper panels) and for Thouless states projected
for particle number $A=24$ and $T=S=0$ (lower panels).
The red dots and red band indicate the minima of
energies, that is, in the upper and lower panels they indicate
solutions of the HFB and VAP equations, respectively. We see that in
all cases the minima of energies appear at $\varphi=\tfrac{\pi}{4}$,
that is, for time-even Thouless states.

For the unprojected states, for $x<0$ the minima stay at $\alpha=\pi$
(purely isovector pairs) and then for $x>0$ they flip over to $\alpha=0$
(purely isoscalar pairs). At $x=0$, the HFB energy is entirely independent
of $\alpha$, so that states with any isovector-isoscalar pair mixing
are exactly degenerate. Our HFB results confirm the observations of
Ref.~\cite{(Eng97)} that the unprojected mean-field states
do not exhibit isovector-isoscalar pairing mixing.
However, as we see in the lower panels of Fig.~\ref{Fig:maps}, our
VAP states do exhibit such a mixing. Indeed, even a
small departure from the pure isovector or isoscalar interaction
moves the VAP solutions away from the unmixed states characterized by $\alpha=0$ or
$\alpha=\pi$. As expected, at $x=0$ the VAP solution appears at
$\alpha=\tfrac{1}{2}\pi$, so that the pairs are then maximally
mixed.

\begin{figure}[h]
\begin{center}
\includegraphics[width=\textwidth]{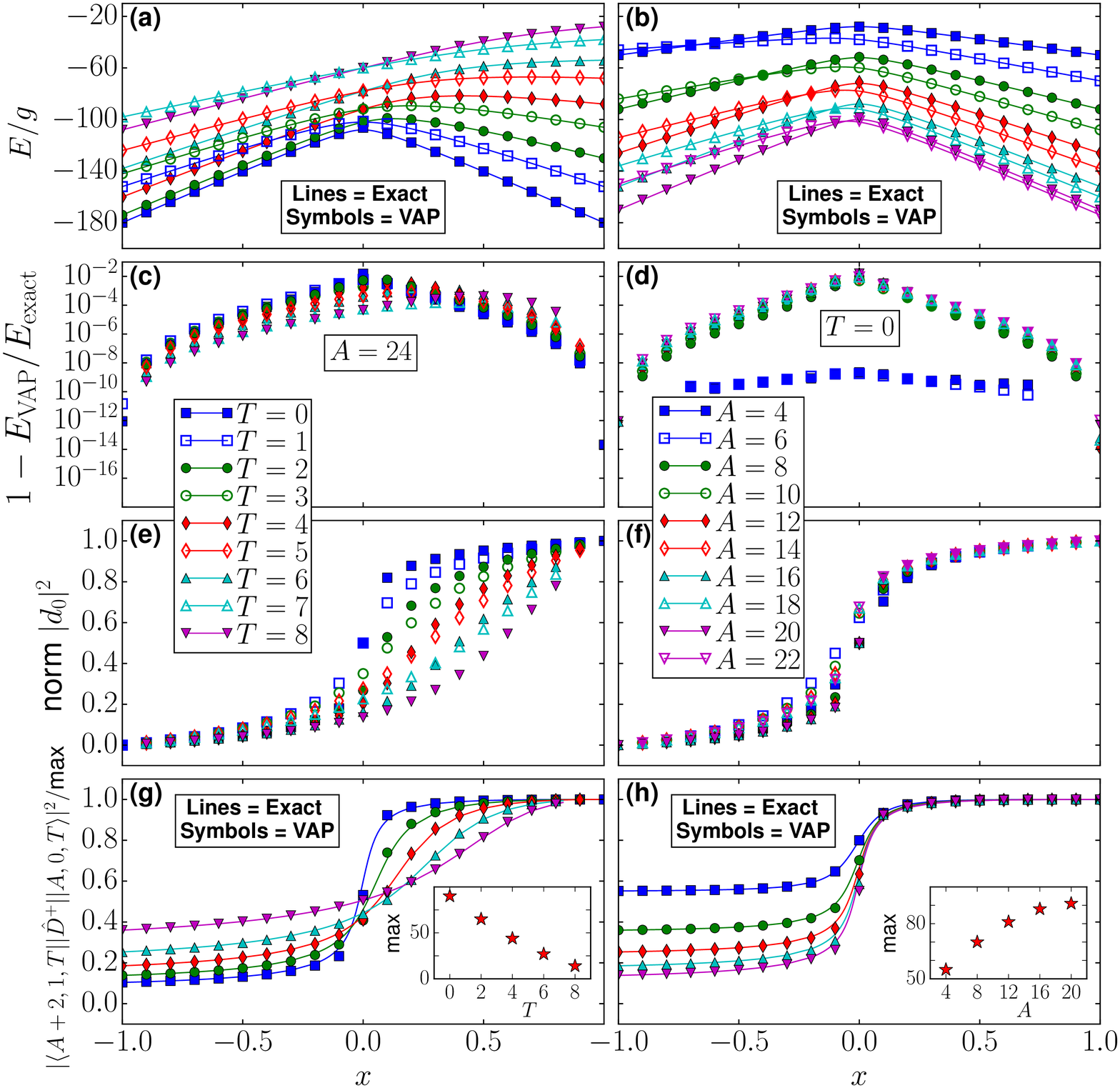}
\end{center}
\caption{Top panels: Energies in units of $g$. Upper middle panels:
Relative errors of the VAP energies shown in the logarithmic scale.
Lower middle panels: Norms of the isoscalar Thouless pairs $|d_0|^2$.
Bottom panels: Deuteron-transfer matrix elements. In the top and bottom
panels, results obtained within the VAP method (symbols) are compared
with those corresponding to the exact solutions (lines). Left panels
show results obtained for $A=24$, with the isospin increasing from
$T=0$ to 8, and spin $S=0$ or 1 for even and odd $T$, respectively.
Right panels show results obtained for $T=0$, with the particle
number increasing from $A=4$ to 22, and spin $S=0$ or 1 for even and
odd $A/2$, respectively.
}
\label{Fig:energies}
\end{figure}

Let us now discuss the VAP solutions, that is, properties of states
$|AST\rangle$ that are projected on good particle number $A$, spin $S$, and isospin
$T$ with energies minimized over $\alpha$ at
$\varphi=\tfrac{\pi}{4}$. Figure \ref{Fig:energies} summarizes our
VAP results obtained for different isospins (left panels) and
particle numbers (right panels)\footnote{We plot VAP results only
for projected states {$|AST\rangle$} that have numerically
significant norms}.
As one can see in the top panels of the figure, when plotted on a
linear scale, the VAP energies (symbols) are indistinguishable from
the exact values (lines).

Only by plotting energy differences on a logarithmic scale
(upper middle panels) can one appreciate the fact that at $x=0$ the
VAP energies are precise up to 1.5\%, and that with growing $|x|$ their
precision rapidly improves by many orders of magnitude. In the limits
of $x=-1$ or $x=+1$, the Thouless pairs correspond to $S=0$ or $T=0$,
respectively, and thus it is enough to restore either the isospin or
spin symmetry. Then, as already noted in Ref.~\cite{(Dob98a)}, the
VAP results become exact. Here we have shown that even in a more
realistic case of mixed pairing the VAP results constitute an
excellent approximation to the exact ones. We also note that for the
multi-level $T=1$ pairing model very good results were obtained in
Ref.~\cite{(Che78)} by using the GCM mixing of the isospin-restored
HFB states, with pairing gaps used as generator coordinates. In light
of our findings, one can interpret such a GCM approach as leading to
analogous solutions to those that we obtain by employing the
full VAP method. Finally, as one can see in
Fig.~\ref{Fig:energies}(d), the VAP results obtained for $A=4$ and 6
are for all values of $x$ exact, that is, precise up to the numerical
accuracy, see discussion below.

The lower middle panels of Fig.~\ref{Fig:energies} show norms of the VAP Thouless isoscalar
pairs defined as $|d_0|^2=\cos^2(\tfrac{1}{2}\alpha_{\text{min}})$,
cf.~Eqs.~(\ref{eq:thoupair}) and (\ref{eq:parametrization}). Again we
see that for arbitrary strengths of the isoscalar vs.~isovector
interactions, the VAP isoscalar and isovector pairs do coexist. As
illustrated in Fig.~\ref{Fig:energies}(e), at a given interaction
strength $x>0$, the role of the isoscalar pairs gradually decreases
with isospin $T$, however, even for high values of $T$ their
contributions are still significant.

A possible experimental evidence of coexistence between isoscalar and
isovector pairing can be the observation and analysis of deuteron
transfer reaction~\cite{(Fro71),(Isa05),(Isa18)}, which depends on the reduced
isoscalar-pair transfer matrix element
$\langle{}A+2,S=1,T||D^+||A,S=0,T\rangle$. In the bottom panels of
Fig.~\ref{Fig:energies}, we compare the VAP and exact values of these
matrix elements calculated in the SO(8) model. Here we show results
normalized by the maximum values obtained at $x=1$, whereas the insets
show these maximum values plotted in the absolute scale. Again we see
that the VAP results (symbols) are indistinguishable from the exact
values (lines).

On the one hand, the relative deuteron transfer amplitudes increase
with the strength of the isoscalar interaction, but this increase is
fairly gradual, especially at higher isospins. On the other hand,
absolute values of these amplitudes gradually decrease with the
isospin. So, as expected, the observation of the strong deuteron
transfer is most likely in $N=Z$ nuclei, however, for $N\neq{}Z$, the
effect does not abruptly disappear. The SO(8) model is too
simplistic to draw quantitative conclusions and an analysis
performed in a realistic shell-structure setting is very much
required.

The fact that the projected pair condensates properly describe
isovector and isoscalar pairing correlations can be best seen by
analyzing the simplest case of four particles. Then, the
particle-number projected condensate is given by the square of
the Thouless pair (\ref{eq:thoupair}), that is, by $|\Phi_4\rangle =
{\cal N}'(\hat{Z}^+)^2|0\rangle$. However, the square of the Thouless
pair is equal to a linear combination of five quartets:
$(P^+P^+)^{(00)}$, $(D^+D^+)^{(00)}$, $(P^+D^+)^{(11)}$,
$(P^+P^+)^{(02)}$, and $(D^+D^+)^{(20)}$, where superscripts
$(ST)$ denote values of the total spin $S$ and isospin $T$.
Restoration of the spin and isospin symmetries corresponds in this
case to keeping only the first two, scalar-isoscalar quartets, and
removing the other three. Thus the symmetry-projected
$|AST\rangle=|400\rangle$ state becomes an exact linear combination
of the two basic quartets~\cite{(San15)}. Similarly, the
symmetry-restored state $|610\rangle$ corresponds to an exact linear
combination of these same two basic quartets supplemented by one
vector-isoscalar pair $D^+$ (\ref{eq:pair-isoscalar}). As a result, the
$A=4$ and 6 VAP solutions shown in Fig.~\ref{Fig:energies}(d) are
identical to the exact ones. For larger particle numbers or isospins,
the success of the VAP approach in describing the pair condensation
relies on the fact that it properly accounts for the main components
of the wave functions being given by the two basic scalar-isoscalar
quartets.

We note here that the pn pairing models have already been
intensely analyzed within the quartet-condensation models, see
Refs.~\cite{(San15),(San18)} and references quoted therein. These
references have often compared results with those obtained within
pair-condensation models employed without full VAP symmetry restorations
and concluded that the latter ones were inferior. At variance with
those conclusions, our results show that the obtained inferiority was
not related to the pair-condensate approximation itself, but rather to
the lack of the full VAP symmetry restoration. We stress that
approaches aiming to mix the isovector and isoscalar pairing necessarily
mix the isovector ($T=1$) and vector ($S=1$ or $J=1$) pairs, and thus
a {\em simultaneous} restoration of isospin and angular momentum is
mandatory~\cite{(Rom19)}.

It is now obvious that the effects of the pn-pair condensation
should be analyzed in a more sophisticated setting than that
envisaged up to now. Within a mean-field approach, it appears that
only by performing the VAP calculations one can fully
account for a subtle balance between the isovector and isoscalar
pairing correlations.

Methods to obtain full VAP results for realistic density functionals
have already been formulated~\cite{(She00c)}, and implemented~\cite{(Sto07e)}, in the
simplest case of the particle-number restoration.
When combined with the full restoration of rotational and isospin
symmetries, which were implemented without pairing in
Ref.~\cite{(Sat12a)}, and with the seven-dimensional symmetry
restoration implemented in this Letter, a complete approach is
possible and is now being constructed.

For the Coulomb isospin mixing included together with pairing, a
reduction of the three-dimensional isospin restoration to one
dimension is not possible. Moreover, the former will anyhow be
required if the isocranking
technology~\cite{(Sat01d),(Sat01e),(Sat13e),(She14b)} is used to
control the isospin degree of freedom. However, for axial nuclei, a
one-dimensional integration suffices, so altogether we are
then faced with five-dimensional integrals -- which is a fully manageable
task. Before attacking the full VAP approach, the results
of this Letter indicate that a restricted minimization of the
projected energies with respect to relative amplitudes of the
isovector and isoscalar pairs could be a viable simplifying option.

The possibility of implementing such a methodology in a realistic
setting of microscopic density functionals crucially depends on
developing functionals based on density-independent
generators~\cite{(Ben17a)} with controlled isoscalar vs.~isovector
pairing strengths. We have already implemented the second aspect by
adding to the inventory of generators terms separable in the pairing
channel, cf.~Refs.~\cite{(Dug04),(Tia09d),(Nik10),(Ves12)}. The work
towards obtaining functionals suitable for the full VAP treatment of
the pn pairing is now being intensely pursued.

In conclusion, within a simple SO(8) pairing model, we have shown that
the symmetry-projected condensates of mixed isovector and isoscalar pairs very accurately describe
properties of the exact solutions, including the coexistence of the
isovector and isoscalar pairing. Lack of symmetry restoration
thus explains the limited success in describing such a coexistence in the
standard mean-field approaches to date. Symmetry restoration is also key to
reconciling the pair-condensation and quartet-condensation pictures
of paired systems. Our study suggests that further work on properties
of the proton-neutron nuclear pairing should be, and can be, carried
out within the variation-after-projection approach to mean-field
pairing methods.

\bigskip\noindent
This work was partially supported by the STFC Grants No.~ST/M006433/1
and No.~ST/P003885/1.

\newpage

\bibliographystyle{elsarticle-num}

\end{document}